\documentclass{article}

\usepackage{microtype}
\usepackage{graphicx}
\usepackage{subcaption}
\usepackage{booktabs} 

\usepackage{bm}
\usepackage{graphicx}
\usepackage{authblk}
\usepackage{tikz}
\usetikzlibrary{positioning, shapes.geometric, calc}
\usepackage{xcolor}
\usepackage{tcolorbox} 
\usepackage{hyperref}

\def\reals{{\mathbb R}}

\newcommand{\es}{{\mathcal{E}}}
\newcommand{\scale}{{\mathcal{S}}}
\newcommand{\psol}{{P_{\text{solve}}}}

\newcommand{\sparseIM}{{\mathcal{I}^{\scale, \Phi}_{\text{sparse}}}}

\usepackage[accepted]{icml2025}

\usepackage{amsmath}
\usepackage{amssymb}
\usepackage{mathtools}
\usepackage{amsthm}

\usepackage[capitalize,noabbrev]{cleveref}

\theoremstyle{plain}
\newtheorem{theorem}{Theorem}[section]

\newtheorem{lemma}[theorem]{Lemma}

\theoremstyle{definition}

\theoremstyle{remark}

\usepackage[textsize=tiny]{todonotes}

\icmltitlerunning{Energy Scale Degradation in Sparse Quantum Solvers}

\begin{document}

\twocolumn[
\icmltitle{Energy Scale Degradation in Sparse Quantum Solvers:\\ A Barrier to Quantum Utility}



\icmlsetsymbol{equal}{*}

\begin{icmlauthorlist}
\icmlauthor{Thang N. Dinh}{vcu}
\icmlauthor{Cao P. Cong}{vcu}
\end{icmlauthorlist}

\icmlaffiliation{vcu}{Department of Computer Science, Virginia Commonwealth University, Richmond, Virginia, USA, 23284}

\icmlcorrespondingauthor{Thang N. Dinh}{tndinh@vcu.edu}

\icmlkeywords{Machine Learning, ICML}

\vskip 0.3in
]




\begin{abstract}
Quantum computing offers a promising route for tackling hard optimization problems by encoding them as Ising models.  However, sparse qubit connectivity requires the use of minor-embedding, mapping logical qubits onto chains of physical qubits, which necessitates stronger intra-chain coupling to maintain consistency. This elevated coupling strength
forces a rescaling of the Hamiltonian due to hardware-imposed limits on the allowable ranges of coupling strengths, reducing the energy gaps between competing states, 
thus, degrading the solver's performance. 
Here, we introduce a theoretical model that quantifies this degradation.  We show that as the connectivity degree increases, the effective temperature rises as a polynomial function, resulting in a success probability that decays exponentially.  Our analysis further establishes worst-case bounds on the energy scale degradation based on the inverse conductance of chain subgraphs, revealing two most important drivers of chain strength, \textit{chain volume} and \textit{chain connectivity}.  Our findings indicate that achieving quantum advantage is inherently challenging. Experiments on D-Wave quantum annealers validate these findings, highlighting the need for hardware with improved connectivity and optimized scale-aware embedding algorithms.
\end{abstract}

\section{Introduction}
Quantum computing, particularly through quantum annealing, promises new routes to obtain high-quality solutions for hard optimization problems. Many of these problems can be efficiently mapped to Ising models \cite{lucas2014ising}, offering a promising path to quantum advantage \cite{Preskill_2018}. Recent milestones, including Google's demonstration of quantum supremacy \cite{arute2019quantum} and advances in quantum annealing hardware like D-Wave's systems \cite{Boothby2020}, bring us closer to practical quantum utility. 

Despite these advances, fundamental physical limitations in current quantum architectures pose significant barriers to achieving quantum utility in solving optimization problems.
A primary challenge is the limited qubit connectivity of physical devices. For instance,  state-of-the-art quantum annealers like D-Wave's Advantage feature 5000+ qubits but only 15 connections per qubit \cite{Boothby2020}.  This sparsity requires the use of minor embedding techniques \cite{choi2008minor,choi2011minor} to map logical qubits onto chains of physical qubits. While such embedding techniques enable fully connected problem mapping, they introduce a trade-off between chain consistency and energy preservation, often resulting in degraded performance as problem size increases.

Moreover, hardware-imposed limits on the ranges of coupling strengths and biases require that the problem Hamiltonian be rescaled when strong intra-chain couplings, termed \textit{chain strength}, are applied. This rescaling compresses the energy gaps between competing states, further impairing the annealer's ability to distinguish the ground state from excited states. 
Previous work has mainly addressed chain length and embedding optimization \cite{choi2011minor, noise_SK, fangChainStrength2020}, leaving a gap in our understanding of \textit{how sparse connectivity fundamentally affects system energy scales and solution quality}.

In this work, we develop a mathematical model that quantifies this degradation with a focus on rescaling effects. Specifically, we show that as the connectivity degree of the problem Hamiltonian increases, the effective energy scale decreases polynomially, resulting in an exponential reduction in success probability. This degradation is not merely a technical limitation but rather a fundamental consequence of sparse connectivity, representing a critical obstacle to achieving quantum advantage.

Our work makes several key contributions to understanding and addressing these challenges:
\begin{itemize}
    \item We develop a theoretical model quantifying energy-scale reduction in sparse quantum solvers under \textit{rescaling effects}, revealing an approximately $O(\sqrt{\Delta})$ degradation in effective energy scales, where $\Delta$ is the maximum degree of spin interactions.
    \item Contrary to the common belief that long chain is the main driver for large chain strengths, we show that the degree of spin interactions $\Delta$ is the main culprit.
    \item We establish worst-case bounds on effective energy scales as functions of both inverse conductance of chain subgraphs and chain lengths. This further reveals two main drivers for setting chain strength value: the degree of interaction, aka \textit{the chain volume} and \emph{the chain connectivity}. In general, chains with large volumes require larger chain strength, while chains with better intra-chain connectivity require lower chain strength.
    \item We validate our theoretical results through experiments on D-Wave quantum annealers and discuss pathways towards quantum advantages through hardware with improved connectivity and more effective scale-aware minor-embedding algorithms.
\end{itemize}

\paragraph{Related work.} To bridge the gap between arbitrary problem structures and hardware constraints, minor embedding maps logical qubits onto chains of physical qubits \cite{choi2008minor, choi2011minor}. Several efficient algorithms have been developed for this NP-complete problem: Cai et al. \cite{embedding} created a heuristic for sparse graphs with hundreds of vertices, Klymko et al. \cite{klymko_adiabatic_2012} provided a polynomial-time algorithm for generating clique minors, and Boothby et al. \cite{Boothby2016} designed an efficient $O(N^3)$ algorithm for finding maximum native clique minors in Chimera graphs.

A critical challenge in embedding is determining appropriate chain strength—the coupling between physical qubits representing the same logical qubit. This strength typically increases with problem graph degree \cite{choi_minor-embedding_2011, noise_SK}, potentially degrading solution quality. Fang et al. \cite{fang2020minimizing} derived tighter bounds for chain strength that can be computed in $O(D2^L)$, showing that optimal chain strength varies significantly with graph structure. Recent work by Gilbert et al. \cite{gilbert2023} and Pelofske \cite{pelofske20234} demonstrates that the topology used for encoding logical qubits (chains vs. cliques) also significantly impacts performance.

Venturelli et al. \cite{noise_SK} empirically showed chain strength scales as O($\sqrt{N}$) for fully-connected spin glasses, while Raymond et al. \cite{raymond2020improving} proposed that chain strength should be tuned proportionally to the square root of the problem size. Choi \cite{choi2020effects} demonstrated that rescaling coupler strength in the problem Hamiltonian proportionally rescales the minimum spectral gap, directly affecting annealing performance. Despite these advances, Koenz et al. \cite{koenz_AnnealingEmbeddingOverhead_2021} showed that current embedding schemes still introduce substantial overhead that scales unfavorably with problem size, highlighting the need for novel approaches to quantum annealing.

Our work advances these results, providing rigorous theoretical proof that chain strength scaling generalizes to O($\sqrt{\Delta}$) for problems with maximum degree $\Delta$. This extension is important since real-world optimization problems rarely require complete connectivity but often contain high-degree hubs. We demonstrate that the dominant factors determining chain strength requirements are chain volume and chain connectivity, not simply chain length as previously assumed, offering new insights into the energy scale degradation that limits quantum advantage.

\paragraph{Organization.} The remainder of this paper is organized as follows. Section~\ref{sec:prelim} establishes the theoretical foundations, including Ising models, minor embedding, and abstract Ising machines. Section~\ref{sec:sparse_ising} provides modeling and analysis of sparse Ising machines in solving optimization problems, covering chain consistency and energy scaling relationships. Section~\ref{sec:high_degree} presents our main theoretical results on energy scale degradation, proving fundamental bounds and limitations. 
We derive an upper bound on the chain-strength through the conductance of individual chain in Section~\ref{section:conductance}
Section~\ref{sec:conclusion} discusses implications and future research directions.

\section{Preliminaries}
\label{sec:prelim}

\subsection{Ising Hamiltonian Formulation}

Combinatorial optimization problems can be efficiently encoded as Ising Hamiltonians, providing a unified framework for quantum optimization approaches \cite{lucas2014ising}. In the Ising model, binary variables are represented by interacting spins located at vertices of a graph $G = (V, E)$. Each spin $s_i$ associated with vertex $i \in V$ takes a value in $\{-1,+1\}$, and a configuration of all spins forms a system state $s \in \{-1,+1\}^n$, where $n = |V|$. The energy of a spin configuration is determined by the problem Hamiltonian:

\begin{equation}
   H_p(s) = \sum_{i \in V} h_i s_i + \sum_{(i,j) \in E} J_{ij} s_i s_j
\end{equation}

where $h_i \in \reals$ represents the local field acting on vertex $i$, and $J_{ij} \in \reals$ denotes the coupling strength between vertices $i$ and $j$. The ground state of this system—the spin configuration that minimizes $H_p(s)$—corresponds to the optimal solution of the original optimization problem.

\subsection{Adiabatic Quantum Computing and Quantum Annealing}

Adiabatic Quantum Computing (AQC) leverages the adiabatic theorem \cite{born_beweis_1928, moritaMathQA2008} to solve optimization problems. This theorem states that a quantum system initially in its ground state will remain in the ground state if the Hamiltonian changes sufficiently slowly.

Quantum annealing exploits this principle by evolving a quantum system from an initial simple Hamiltonian $H_d$ (driver Hamiltonian) to the problem Hamiltonian $H_p$ according to:

\begin{equation}
   H(t) = A(t)H_d + B(t)H_p
\end{equation}

where $A(t)$ decreases from 1 to 0 and $B(t)$ increases from 0 to 1 over annealing time $T$. The initial Hamiltonian $H_d$ typically implements quantum tunneling through transverse fields, while the final Hamiltonian $H_p$ encodes the optimization problem \cite{albashAdiabaticQC2018}.

Quantum annealing (QA) can be viewed as a relaxation of AQC, where the annealing schedule is determined heuristically without guaranteeing strict adiabatic conditions \cite{hauke_perspectives_2019}. This makes QA a heuristic optimization approach that maintains a non-zero probability of finding the ground state.

D-Wave Systems has developed the largest-scale quantum annealers, with their latest generations containing over 5000 qubits \cite{Boothby2020}. Other initiatives include NEC's parametron qubits and Qilimanjaro's high-coherence processors \cite{qilimanjaro}.

Current quantum annealers feature limited connectivity patterns between qubits. D-Wave's earlier systems used the Chimera topology, where each qubit connects to six others, while their newer Advantage systems employ the Pegasus topology, with each qubit connecting to fifteen others, enabling implementation of denser problem graphs \cite{pegasus}. More recently, the Zephyr topology has been introduced, which further enhances connectivity by allowing each qubit to connect to twenty others. This increased connectivity facilitates even denser problem embeddings, though it comes with additional engineering challenges \cite{dwave2021zephyr}.

\subsection{Optimization on Sparse Quantum Solvers}

The process of solving optimization problems on sparse quantum annealers involves several key steps to bridge the gap between arbitrary problem structures and hardware constraints.

\paragraph{Minor-embedding.} Let $G_p = (V_p, E_p)$ be the logical problem graph underlined the problem Ising Hamiltonian $H_p$.  Denote by $G_{hw} = (V_{hw}, E_{hw})$ 
 the physical hardware graph, representing the qubit connectivity. Due to the lack of connectivity in $G_{hw}$, we need to map $G_p$ to $G_{hw}$
through a mapping $\phi: G_p \rightarrow G_{hw}$ \cite{choi2008minor, choi2011minor} in which each logical qubit $i \in V_p$ will be mapped to a subset of qubit(s) in $G_{hw}$ denoted by $C_i$. The subset of physical spins $C_i$, inducing a connected subgraph in $G_{hw}$, is called a \textit{chain} and the number of spins in $C_i$ is referred as \textit{chain length}.
To ensure these chains function as single logical units, strong ferromagnetic couplings (negative values in the Ising model) are applied between physical qubits within each chain. The magnitude of these couplings is called the \textit{chain strength} (denoted by $\lambda > 0$) and must be sufficiently large to maintain \textit{chain consistency} -- the state where all physical qubits within a chain have identical spin values.

We denote the \textit{embedded Hamiltonian}, parameterized by the chain strength $\lambda$, on the hardware graph as
\begin{equation}
  H_e^{\lambda}(s) = \sum_{u \in V_{hw}} h_u s_u + \sum_{(u,v) \in E_{hw}} J_{uv} s_u s_v
\end{equation}

Let $E(C_i, C_j)$ be the set of \textit{inter-chain connections} from the qubits in $C_i$ to those in $C_j$ within $E_{hw}$. The sum of the strengths on those edges must equal the original logical coupling, i.e., $$\sum_{(u,v) \in E(C_i, C_j)} J_{uv} = J_{ij}$$ to preserve the energy landscape. Typically, these coupling strengths are distributed equally as $J_{uv} = J_{ij}/|E(C_i, C_j)|$. For \textit{intra-chain connections} (edges within the same chain), a strong negative coupling $J_{uv} = -\lambda$ is applied to maintain chain consistency. Similarly, the linear bias $h_i$ is distributed, often equally, among the $h_u$ for $u$ in chain $C_i$.

\paragraph{Chain consistency handling.} After embedding, the quantum annealer solves this embedded problem according to its physical constraints and returns samples from the embedded solution space. However, due to finite temperature effects and other noise sources, physical qubits within a chain may not always maintain consistent values, resulting in \textit{chain breaks}.   Returned solutions without chain consistency can be either filtered out or amended using common approaches such as majority voting (assigning the logical value based on the majority state of physical qubits) or minimum energy (assigning the logical value that minimizes the energy).

\subsection{Increased Chain Strength And Energy Scale Compression}
Quantum annealing hardware imposes physical limitations on the allowable ranges for coupling strengths and biases. 
 For example, D-Wave systems use extended ranges of $J_{\text{range}} = [-2, 1]$ for couplings and $h_{\text{range}} = [-4, 4]$ for local fields, though these ranges may vary across different quantum annealing platforms  \cite{Boothby2020}.
 This necessitates a rescaling of the Hamiltonian to fit the ranges.

When embedding a problem, the introduction of intra-chain couplings ($-\lambda$) leads to an inherent trade-off. Increasing the chain strength $\lambda$ enhances chain consistency but simultaneously reduces the scale factor $\scale$. As $\scale$ decreases, the coupling strengths and biases of the original problem are proportionally reduced, diminishing the energy gaps between competing states. Consequently, the annealer's ability to distinguish the ground state from excited states is compromised.

It has been shown that the magnitude of the minimum chain strength increases with the degree of the graph of $H_p$~\cite{choi_minor-embedding_2011, noise_SK,paintshop}, which can create distortions in the resulting Hamiltonian due to noise. 

This fundamental tension—requiring strong chains for consistency while simultaneously preserving the problem's energy scale—creates one of the central challenges in quantum annealing implementations. In subsequent sections, we develop a theoretical model quantifying this energy scale reduction and its impact on solution success probability, providing a framework for understanding the fundamental limitations of sparse quantum solvers.

\begin{figure*}[th!]
    \centering
    \begin{subfigure}{.20\linewidth}
        \label{fig:testa}
        \centering
        \includegraphics[width=\linewidth]{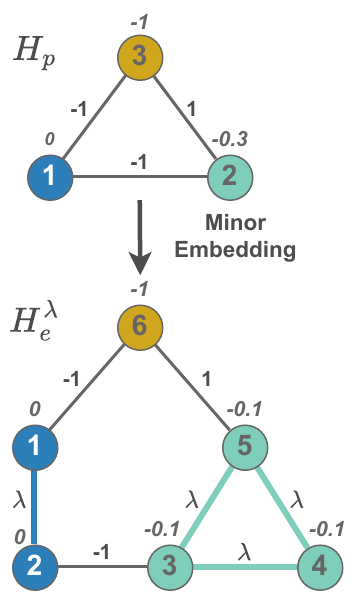}
        \caption{\small \centering The problem and embedded Hamiltonians}
    \end{subfigure}
    \begin{subfigure}{.25\linewidth}
        \label{fig:testb}
        \centering
        \includegraphics[width=\linewidth]{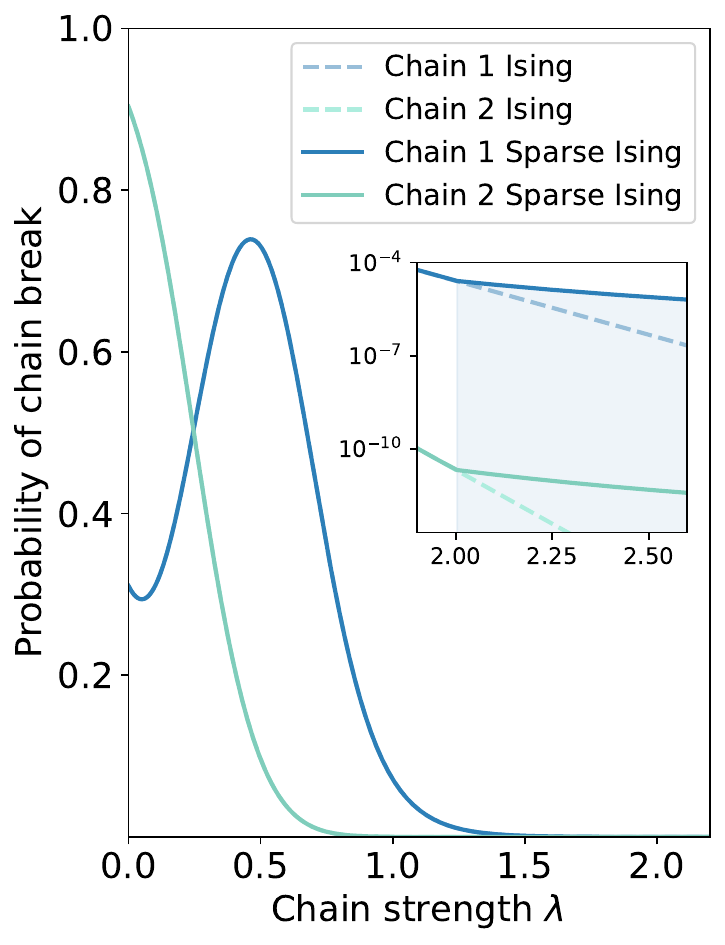}
        \caption{\small \centering Individual chain break on Ising models}
    \end{subfigure}
    \begin{subfigure}{.26\linewidth}
        \label{fig:testc}
        \centering
        \includegraphics[width=\linewidth]{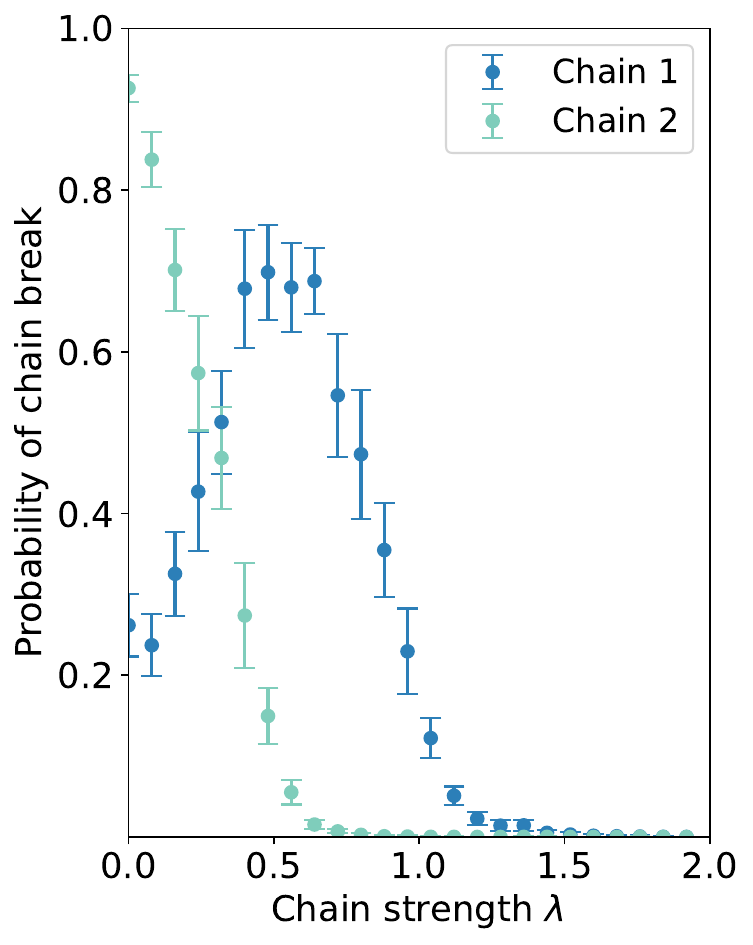}
        \caption{\small \centering Individual chain break on D-Wave's quantum annealer}
    \end{subfigure}
    \begin{subfigure}{.26\linewidth}
        \label{fig:testd}
        \centering
        \includegraphics[width=\linewidth]{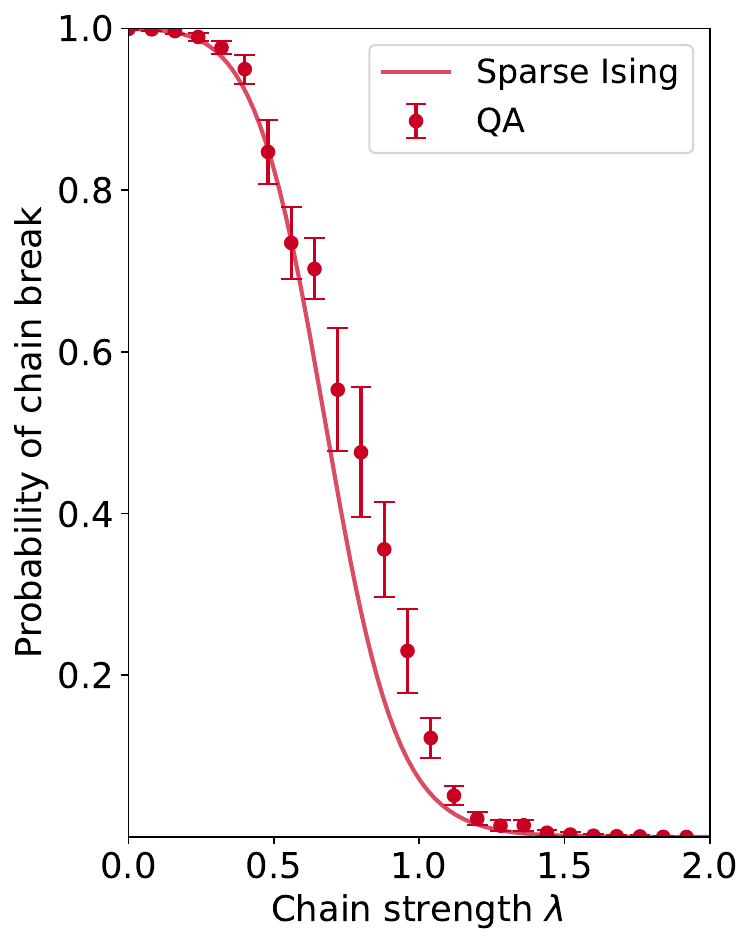}
        \caption{\small \centering Chain break probabilities ($1~-~P_{cc}$)}
    \end{subfigure}
    \caption{
    Energy rescaling effects on chain consistency in sparse quantum solvers. \textbf{(a)} Illustration of the problem Ising Hamiltonian $H_p$ and its hardware embedding $H_e^\lambda$. The values on nodes and edges represent linear biases and coupling strengths, respectively. The embedding maps logical variables to physical chains: \textcolor[rgb]{0.172, 0.498, 0.722}{chain 1 ($C_1$)} consisting of physical spins 1 and 2, and \textcolor[rgb]{0.498, 0.804, 0.733}{chain 2 ($C_2$)} consisting of physical spins 3, 4, and 5. 
    \textbf{(b)} Non-monotonic chain break probability as a function of chain strength $\lambda$ at inverse temperature $\beta = 4$. Counterintuitively, as $\lambda$ increases from 0.1 to 0.5, the break probability for \textcolor[rgb]{0.172, 0.498, 0.722}{chain 1} increases while decreasing for \textcolor[rgb]{0.498, 0.804, 0.733}{chain 2}. The inset shows how energy rescaling in the sparse Ising model (solid lines) elevates chain break probabilities compared to the standard Ising model (dashed lines) for $\lambda > 2$.
    \textbf{(c)} Experimental validation using D-Wave \texttt{Advantage\_system4.1} quantum annealer (20 runs, 400 samples each), showing consistent non-monotonic behavior for \textcolor[rgb]{0.172, 0.498, 0.722}{chain 1}.
    \textbf{(d)} Theoretical predictions from the sparse Ising model closely match experimental quantum annealing results.
    }
    \label{fig:test}
\end{figure*}

\section{Sparse Ising Machine With Energy Rescaling}
\label{sec:sparse_ising}
\subsection{Modeling Sparse Ising Machine with Rescaling}
\label{sec:ising_rescaling}

We define an abstract Ising Machine model for sparse quantum solvers that captures the essential elements of embedding and energy rescaling. This model, denoted by $\sparseIM$, is characterized by an inverse temperature $\beta$ and a minor-embedding algorithm $\Phi$ to embed each problem Ising Hamiltonian $H_p$ to the hardware graph $G_{hw} = (V_{hw}, E_{hw})$ 
representing the physical qubits and their connectivity.

\paragraph{Problem Embedding.} Given an Ising Hamiltonian problem $H_p$ defined on a logical graph $G_p = (V_p, E_p)$, a minor-embedding algorithm $\Phi$ maps $H_p$ to the hardware graph $G_{hw}$, creating an embedded Ising Hamiltonian $H_{e}^{\lambda}$ parameterized by the chain strength $\lambda$. The minor-embedding algorithm $\Phi$ also returns the mapping $\phi$ that associates each logical spin $i \in H_p$ to a connected subgraph, called chain, $C_i$ in the hardware graph. That is
\[
(H_{e}^{\lambda}, \phi) = \Phi(H_p, G_{hw} ).
\]
The embedded Hamiltonian has the form:
\begin{equation}
  H_e^{\lambda}(s) = \sum_{u \in V_{hw}} h_u s_u + \sum_{(u,v) \in E_{hw}} J_{uv} s_u s_v
\end{equation}
where the intra-chain couplings are set to $J_{uv} = -\lambda$ for $(u,v) \in E(C_i)$ to encourage consistent spin values within chains, and the inter-chain couplings preserve the logical problem structure.

\paragraph{Energy Scaling.} Due to hardware constraints on the allowable ranges for coupling strengths and biases, a scaling function $\scale$ is applied to the embedded Hamiltonian. This scaling function is defined as:
\begin{equation*}
\scale(H) = \max\left\{\frac{\max\{h_i\}}{\mathcal{M}_{\max}}, \frac{\min\{h_i\}}{\mathcal{M}_{\min}}, \frac{\max\{J_{ij}\}}{\mathcal{J}_{\max}}, \frac{\min\{J_{ij}\}}{\mathcal{J}_{\min}}\right\}
\end{equation*}
where $\mathcal{M}_{\max}$, $\mathcal{M}_{\min}$, $\mathcal{J}_{\max}$, and $\mathcal{J}_{\min}$ represent the hardware-imposed bounds on the magnitudes of biases and couplings. For convenience, we define \[
\es_{\lambda} = \scale(H^{\lambda}_e)\] to shorten the notation when context is clear. The actual Hamiltonian solved on the Ising machine is $ H_e^{\lambda}/\es_{\lambda}$, effectively compressing all energy gaps among competing states by a factor of $\scale_{\lambda}$.

\paragraph{Gibbs Sampling.} Our Ising machine samples solutions follow Gibbs distribution as in the original the Ising model \cite{ising1925beitrag}.
For any Ising Hamiltonian $H$ with spin configuration space $\Omega$ and inverse temperature $\beta$, the Gibbs distribution defines the probability of observing a particular configuration $s$:
\begin{equation}
P(s|H,\beta) = \frac{e^{-\beta H(s)}}{Z}
\end{equation}
where $Z = \sum_{s \in \Omega} e^{-\beta H(s)}$ is the partition function.

\textit{Solution Probability.} We define 
 $\psol(H_p, \beta)$
as the probability of observing any ground state (minimum energy configuration) under the Gibbs distribution:
\begin{equation}
\psol(H, \beta) = \sum_{s \in \Omega_{\text{min}}} P(s|H,\beta)
\end{equation}
where $\Omega_{\text{min}} = \{s \in \Omega \mid H(s) = \min_{s' \in \Omega} H(s')\}$ is the set of ground states.

A summary of our sparse Ising machine with energy rescaling model is presented below.

\begin{tcolorbox}[title=Sparse Ising Machine with Energy Rescaling $\sparseIM$, colback=white]
\begin{enumerate}
    \item \textit{Minor-embedding}: Mapping $H_p$ to the sparse Ising machine using the minor-embedding $\phi$: $(H_e^\lambda, \phi) = \Phi(H_p, G_{hw})$.
    \item \textit{Rescaling}: Shrink the embedded Ising by a factor $\es_{\lambda} = \scale(H_e^\lambda)$ to fit the machine limits on the bias ranges.
    \item \textit{Sampling}: Sample the  embedded and rescaled Ising $\es_{\lambda}^{-1} H_e^\lambda$ solutions on the sparse Ising machine following a Gibbs distribution with an inverse temperature $\beta$.    
    \item \textit{Unembedding}: Discarding samples with any chain breaks (chain break configurations). Recover logical spin values and the corresponding energy for the problem Ising $H_p$ for the remaining samples.
\end{enumerate}
\end{tcolorbox}

We continue defining the two important metrics for the model, namely chain consistency probability and solution probability.

\textit{Chain Consistency Probability.}  For a spin configuration $s$ of $H_e^\lambda$, the configuration is said to have \textit{chain consistency} if for every spin $i$ in $H_p$, all the spins within the corresponding chain $C_i$ have the same value. Otherwise, the configuration will have some chain $C_i$ with spins of opposite signs, aka \textit{chain break}.
$P_{\text{cc}}(H_e^{\lambda}, \sparseIM)$, or  $P_{\text{cc}}(\lambda)$ when the context is clear, is the probability of observing configurations with chain consistency

\begin{equation}
P_{\text{cc}}(H_e^{\lambda}, \sparseIM) = \sum_{s \in \Omega_{\text{cc}}} P(s|H_e^{\lambda}/\es_{\lambda},\beta)
\end{equation}

where $\Omega_{\text{cc}} = \{s \in \Omega \mid \forall i, \forall u,v \in C_i, s_u = s_v\}$ represents the set of configurations with chain consistency, i.e., no chain breaks.

\textit{Solution Probability on Sparse Ising Machines.} Finally, we define $P_{\text{solve}}^{\text{sparse}}(H_p, \beta, \lambda)$ as the probability of obtaining a ground state of the original problem Hamiltonian $H_p$ when sampling on $\sparseIM$ when chain strength is set to $\lambda$.

\paragraph{Example.} An illustration of solving Ising Hamiltonian on  Sparse Ising Machine is shown in Figure \ref{fig:test}. We begin with a problem Ising Hamiltonian $H_p = - 0.3 s_2 - s_1 - s_1 s_2 + s_2 s_3 - s_1 s_3$ and its minor-embedding $H_e^\lambda$ on a hardware graph, as shown in Fig.~\ref{fig:test}(a). The embedding maps logical variables to physical chains: chain 1 ($C_1$) consisting of physical spins 1 and 2, and chain 2 ($C_2$) consisting of physical spins 3, 4, and 5.  The minor-embedding is also rescaled to fit the hardware limits on the coupling and biases ranges. 

Fig.~\ref{fig:test}(b) demonstrates a counterintuitive phenomenon: as chain strength $\lambda$ increases from 0.1 to 0.5, the break probability for chain 1 increases while decreasing for chain 2. This non-monotonic behavior illustrates that the relationship between chain strength and consistency is more complex than commonly assumed. The inset further reveals how energy rescaling elevates chain break probabilities compared to standard Ising models when $\lambda > 2$.

Our experimental validation using D-Wave's quantum annealer, shown in Fig.~\ref{fig:test}(c), confirms this non-monotonic behavior for chain 1, providing strong evidence that our theoretical model captures physical behaviors in quantum annealing hardware. Fig.~\ref{fig:test}(d) shows excellent agreement between our theoretical predictions and experimental results across varying chain strengths.
The example empirically demonstrates (1) energy rescaling effects are significant in sparse quantum solvers, (2) chain consistency behaves in complex, sometimes counterintuitive ways with increasing chain strength.

\subsection{Intractability of  Approximating Chain Consistency Probability}
We show that even approximating the chain consistency probability on sparse Ising machines is intractable under typical complexity assumption. We begin by establishing the hardness of approximating spin correlations in the standard Ising model, then use this result to prove the intractability of estimating chain consistency probability in embedded systems.

Let $\beta_c(\Delta)$, where $\beta_c(\Delta)$ is the critical inverse temperature threshold for the ferromagnetic Ising model on graphs with maximum degree $\Delta$. For temperatures below this critical threshold (i.e., when $\beta > \beta_c(\Delta)$), the Ising model exhibits the "non-uniqueness" phase characterized by multiple Gibbs measures.

\begin{lemma}[Hardness of Spin Correlation Approximation]
\label{lemma:spin_correlation}
Let $\Delta \geq 3$ and $\beta > \beta_c(\Delta)$. Approximating the spin correlation $\mathbb{E}[X_u X_v]$ for any edge $(u, v)$ within any constant additive error is $\mathsf{NP}$-hard, assuming $\mathsf{RP} \neq \mathsf{NP}$.
\end{lemma}

Using this lemma, we can now establish the hardness of approximating chain consistency probability in sparse Ising machines:

\begin{theorem}[Intractability of Chain Consistency Approximation]
\label{theorem:chain_consistency}
For a sparse Ising machine $\sparseIM$ with effective inverse temperature $\beta \cdot \es_{\lambda} > \beta_c(\Delta)$, approximating the chain consistency probability $P_{\text{cc}}(H_e^{\lambda}, \sparseIM)$ within any constant additive error is $\mathsf{NP}$-hard, assuming $\mathsf{RP} \neq \mathsf{NP}$.
\end{theorem}

This theorem reveals a fundamental limitation in sparse quantum solvers: as chain strength $\lambda$ increases to maintain consistency, the scaling factor $\es_{\lambda}$ decreases, potentially causing $\beta \cdot \es_{\lambda}$ to approach (or remain above) the critical threshold $\beta_c(\Delta)$ where chain consistency probability becomes computationally intractable to approximate. 

This result explains why selecting optimal chain strength in practice remains challenging—accurately predicting how a given chain strength affects consistency probability is fundamentally hard.

\begin{figure}[t]
    \centering
    \begin{subfigure}{.49\linewidth}
        \centering
        \includegraphics[width=\linewidth]{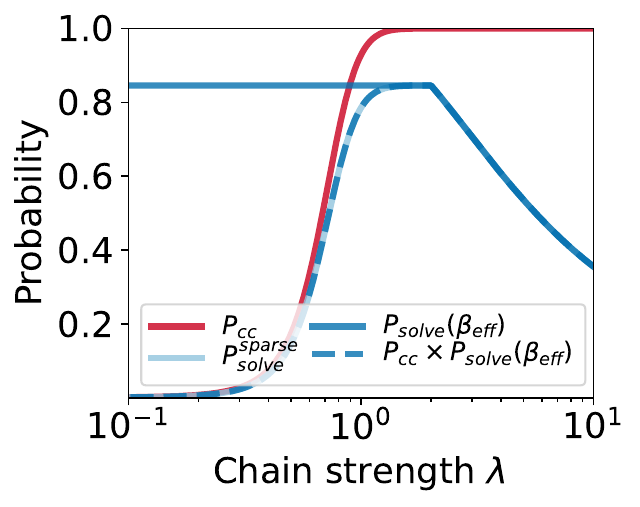}
        \caption{\small \centering Sparse Ising Machine with Energy Rescaling}
    \end{subfigure}
    \begin{subfigure}{.49\linewidth}
        \centering
        \includegraphics[width=\linewidth]{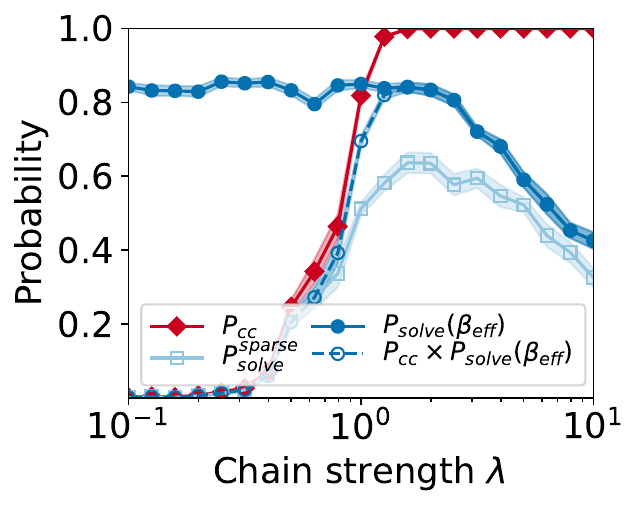}
        \caption{\small \centering D-Wave Quantum Annealer \texttt{Advantage\_system4.1}}
    \end{subfigure}
    \caption{\small
    Chain consistency and solution probability trade-off as a function of chain strength $\lambda$, comparing \textbf{(a)} theoretical sparse Ising model with energy rescaling and \textbf{(b)} experimental results from D-Wave's quantum annealer for the instance in Fig.~\ref{fig:test}. 
    \textcolor[rgb]{0.831, 0.200, 0.298}{$P_{cc}$} (red) represents the chain consistency probability, which generally increases with $\lambda$.  
    \textcolor[rgb]{0.216, 0.553, 0.749}{$\psol(\beta_{\text{eff}})$} (solid blue) shows the solution probability with effective inverse temperature $\beta_{\text{eff}} = \beta/\es_\lambda$, which decreases as $\lambda$ increases due to energy scale compression.
    \textcolor[rgb]{0.655, 0.816, 0.894}{$P_{\text{solve}}^{\text{sparse}}$} (light blue) shows the actual solution probability, while the dashed blue curve shows the theoretical prediction \textcolor[rgb]{0.216, 0.553, 0.749}{$P_{cc} \times \psol(\beta_{\text{eff}})$} according to Theorem~\ref{theorem:solve_probability}.
    Both theoretical and experimental results demonstrate the competing effects of increasing chain consistency versus degrading energy resolution as chain strength increases, with optimal performance achieved at moderate chain strength values.
    }
    
    \label{fig:decomposition}
\end{figure}

\subsection{Chain Strength and Solution Probability Reduction}
\label{sec:energy_deformation}

Having established our abstract Ising machine model $\mathcal{I}^{\scale, \Phi}_{\text{sparse}}$ and the associated scaling effects, we now analyze how the energy landscape deforms with increasing chain strength, focusing specifically on solution probabilities.

Let $\Omega_s$ denote the configuration space of physical spins in the hardware graph $G_{hw}$, and $\Omega_p$ denote the configuration space of logical spins in the problem graph $G_p$. We define $\Omega_{\text{cc}} \subset \Omega_s$ as the subspace of chain-consistent configurations where all physical spins within each chain $C_i$ have identical values. 

\begin{theorem}
\label{theorem:solve_probability}
Consider a problem Hamiltonian $H_p$ and a sparse Ising machine $\sparseIM$ with inverse temperature $\beta$ and rescaling function $\scale$. Let $H_e^\lambda$ be the embedding of $H_p$ on $\sparseIM$ with chain strength $\lambda$ and $\es_\lambda = \scale(H_e^\lambda)$ be the scaling factor. The probability of finding a ground state solution using chain strength $\lambda$ is:
\begin{equation}
\boxed{P_{\text{solve}}^{\text{sparse}}(\beta, \lambda) = P_{\text{cc}}(\lambda) \cdot \psol({\beta_{\text{eff}}(\lambda)})}
\end{equation}
where $P_{\text{cc}}(\lambda)$ is the chain consistency probability, $\beta_{\text{eff}}(\lambda) = \beta/\es_\lambda$ is the effective inverse temperature, and $\psol({\beta_{\text{eff}}(\lambda)})$ is the solution probability of $H_p$ at this effective temperature.
\end{theorem}

\paragraph{Key Factors for Solution Probability.}  
Theorem~\ref{theorem:solve_probability} reveals a fundamental trade-off in sparse quantum solvers by decomposing the solution probability into two competing factors:

\begin{enumerate}
    \item \textit{Chain Consistency Factor $P_{\text{cc}}(\lambda)$}:
    This factor generally increases with chain strength $\lambda$, as stronger intra-chain coupling promotes consistent spin alignment within each chain. Higher chain consistency means more valid solutions in the logical space.
    
    \item \textit{Energy Resolution Factor $\psol(\beta_{\text{eff}}(\lambda))$}:
    This factor captures the probability of finding a ground state given the rescaled energy landscape. As $\lambda$ increases, the scaling factor $\es_\lambda$ also increases, reducing the effective temperature parameter $\beta_{\text{eff}}(\lambda) = \beta/\es_\lambda$ and compressing energy differences between states. This diminishes the solver's ability to distinguish the ground state, causing this factor to decrease with increasing $\lambda$.
\end{enumerate}

The overall solution probability reflects the tension between these competing effects. At low chain strengths, chain breaks dominate and limit success. At high chain strengths, energy scale compression becomes the primary limiting factor.

\paragraph{Example. }  We present the validation of the theoretical result on the Ising Hamiltonians in Fig.~\ref{fig:decomposition}, showing the trade-off between chain consistency and solution probability in sparse Ising solvers. Using the Ising in Fig.~\ref{fig:test}, we compare our theoretical model with experimental results obtained from D-Wave's \texttt{Advantage\_system4.1} quantum annealer. Our experimental measurements were conducted using 20 independent runs with 400 samples each on the D-Wave quantum annealer, with 95\% confidence intervals shown as shaded regions.

We observe that chain consistency probability ($P_{cc}$) generally increases with chain strength $\lambda$, as stronger intra-chain coupling promotes alignment among physical qubits within each chain. Simultaneously, the solution probability for the original problem at effective inverse temperature ($\psol(\beta_{\text{eff}})$) decreases as $\lambda$ increases, due to energy scale compression that reduces the effective temperature parameter $\beta_{\text{eff}} = \beta/\es_\lambda$.

The actual solution probability ($P_{\text{solve}}^{\text{sparse}}$) exhibits non-monotonic behavior with respect to chain strength, peaking around $\lambda \approx 2$. 
 These results empirically demonstrate the existence of an optimal chain strength that balances the competing factors of chain consistency and energy resolution. This optimal point represents the maximum achievable solution quality for a given problem embedding, beyond which further increases in chain strength actually degrade performance despite improving chain consistency.

Theorem~\ref{theorem:solve_probability} states that $P_{\text{solve}}^{\text{sparse}} = P_{cc} \times \psol(\beta_{\text{eff}})$. While not entirely identical, the close agreement between the theoretical prediction (dashed blue) and experimental measurements (light blue) across a range of chain strengths demonstrates the effectiveness of our model in capturing the behaviors of sparse quantum solvers.

\paragraph{Empirical Validation.} Chain consistency and solution probability trade-off as a function of chain strength $\lambda$, comparing \textbf{(a)} theoretical sparse Ising model with energy rescaling and \textbf{(b)} experimental results from D-Wave's quantum annealer for the instance in Fig.~\ref{fig:test}. 
    \textcolor[rgb]{0.831, 0.200, 0.298}{$P_{cc}$} (red) represents the chain consistency probability, which generally increases with $\lambda$.  
    \textcolor[rgb]{0.216, 0.553, 0.749}{$\psol(\beta_{\text{eff}})$} (solid blue) shows the solution probability with effective inverse temperature $\beta_{\text{eff}} = \beta/\es_\lambda$, which decreases as $\lambda$ increases due to energy scale compression.
    \textcolor[rgb]{0.655, 0.816, 0.894}{$P_{\text{solve}}^{\text{sparse}}$} (light blue) shows the actual solution probability, while the dashed blue curve shows the theoretical prediction \textcolor[rgb]{0.216, 0.553, 0.749}{$P_{cc} \times \psol(\beta_{\text{eff}})$} according to Theorem~\ref{theorem:solve_probability}.
    Both theoretical and experimental results demonstrate the competing effects of increasing chain consistency versus degrading energy resolution as chain strength increases, with optimal performance achieved at moderate chain strength values.
    Shaded areas indicate 95\% confidence intervals for the D-Wave \texttt{Advantage\_system4.1} quantum annealer (20 runs $\times$ 400 samples).

\paragraph{Quantifying the Energy Resolution Degradation.}
For sufficiently large chain strength $\lambda$, the scaling factor becomes dominated by the chain coupling, giving $\es_\lambda \approx \lambda/|J_{\min}|$, where $|J_{\min}|$ is the minimum coupling magnitude allowed by the hardware. This yields an effective inverse temperature:
\begin{equation*}
\beta_{\text{eff}}(\lambda) = \frac{\beta|J_{\min}|}{\lambda}
\end{equation*}

From statistical physics, for a system with energy gap $\Delta E$ between the ground state and first excited state, the ground state probability scales approximately as:
\begin{align*}
\psol(\beta_{\text{eff}}) &\approx \frac{e^{-\beta_{\text{eff}}E_0}}{e^{-\beta_{\text{eff}}E_0} + e^{-\beta_{\text{eff}}(E_0+\Delta E)}}\\ 
&= \frac{1}{1 + e^{-\beta_{\text{eff}}\Delta E}} \approx 1 - e^{-\beta_{\text{eff}}\Delta E}
\end{align*}

As $\lambda$ increases, $\beta_{\text{eff}}\Delta E \ll 1$, the solution probability decays as
\begin{align*}
\psol(\beta_{\text{eff}}) &\approx \beta_{\text{eff}}\Delta E = \frac{\beta|J_{\min}|\Delta E}{\lambda}
\end{align*}
More generally, as $\lambda$ increases, the solution probability decays as:
\begin{equation*}
\psol(\beta_{\text{eff}}) \approx e^{-\beta_{\text{eff}}\Delta E} = e^{-\beta|J_{\min}|\Delta E/\lambda}
\end{equation*}

\begin{figure}[h!]
    \centering
    \begin{subfigure}{.49\linewidth}
        \centering
        \includegraphics[width=\linewidth]{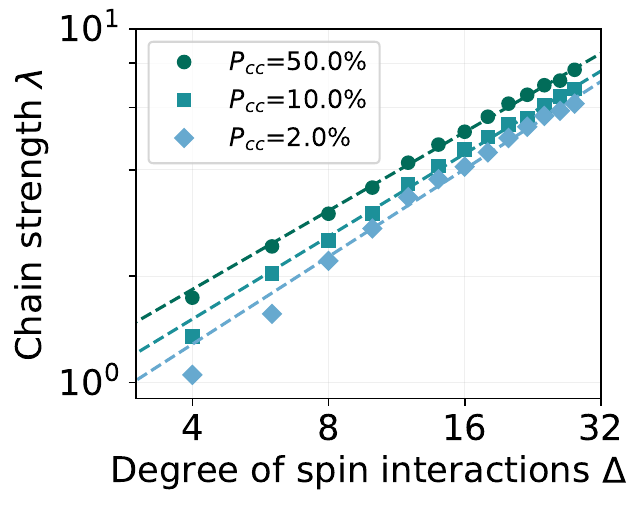}
        \caption{\small \centering Required chain strength $\lambda$, varying $\Delta$\ (log-log scale)}
    \end{subfigure}
    \begin{subfigure}{.475\linewidth}
        \centering
        \includegraphics[width=\linewidth]{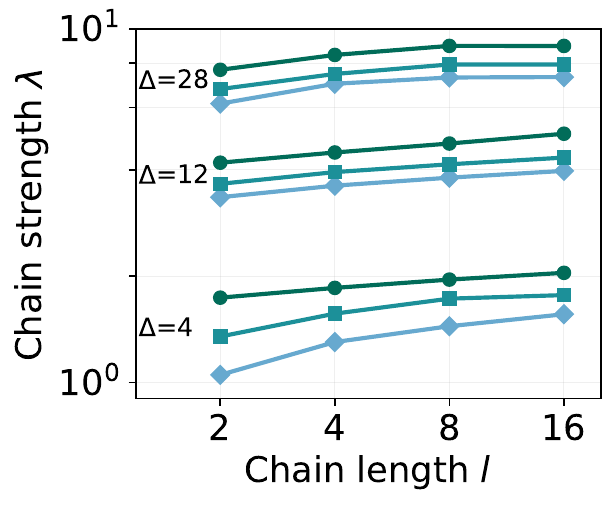}
        \caption{\small \centering Required chain strength $\lambda$, varying $l$\ (log-log scale)}
    \end{subfigure}
    \caption{\small  
    Scaling behavior of required chain strength (log-log scale) to achieve specific chain consistency thresholds (2\%, 10\%, 50\%) on D-Wave \texttt{Advantage\_system4.1}. 
    \textbf{(a)} The dominant effect: Chain strength increases \textit{polynomially} with the degree of spin interactions $\Delta$ (fixed chain length $l=2$), confirming our theoretical prediction of $\lambda \sim O(\sqrt{\Delta})$.
    \textbf{(b)} The secondary effect: Chain strength shows only minor increase with chain length $l$, challenging the common assumption that chain length is the primary driver of required chain strength.
    Dashed lines show linear fits in log-log scale, with slopes indicating scaling exponents. These results demonstrate that the degree of spin interactions, not chain length, is the critical factor determining required chain strength in sparse quantum solvers.
    }
    \label{fig:Delta}
\end{figure}

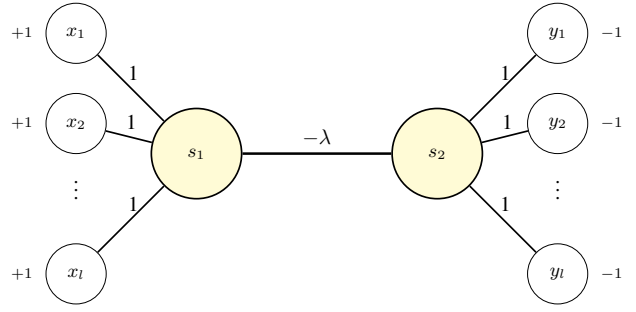
\begin{figure}[t]
    \centering
    \resizebox{\linewidth}{!}{
    \begin{tikzpicture}[
        node/.style={
            draw, 
            circle, 
            minimum size=1cm, 
            inner sep=0pt, 
            font=\small
        },
        central/.style={
            draw,
            circle,
            minimum size=1.5cm,
            fill=yellow!20,
            thick,
            font=\small\bfseries
        },
        edge/.style={
            thick
        },
        bias/.style={
            font=\scriptsize,
            fill=white,
            inner sep=1pt
        }
    ]
        \node[central] (s1) at (0,0) {$s_1$};
        \node[central] (s2) at (4,0) {$s_2$};
    
        \node[node] (x1) at (-2,2) {$x_1$};
        \node[node] (x2) at (-2,0.5) {$x_2$};
        \node[node] (x3) at (-2,-2) {$x_l$};
        \node (dots1) at (-2,-0.5) {$\vdots$};
    
        \node[node] (y1) at (6,2) {$y_1$};
        \node[node] (y2) at (6,0.5) {$y_2$};
        \node[node] (y3) at (6,-2) {$y_l$};
        \node (dots2) at (6,-0.5) {$\vdots$};
    
        \draw[edge, very thick] (s1) -- (s2) node[midway, above] {$-\lambda$};
    
        \foreach \i in {1,2,3} {
            \ifnum \i=1
                \draw[edge] (s1) -- (x1) node[midway, above right, xshift=-5pt] {1};
                \node[bias] at ($(x1) + (-0.9,0)$) {$+1$};
            \fi
            \ifnum \i=2
                \draw[edge] (s1) -- (x2) node[midway, above right, xshift=-5pt] {1};
                \node[bias] at ($(x2) + (-0.9,0)$) {$+1$};
            \fi
            \ifnum \i=3
                \draw[edge] (s1) -- (x3) node[midway, above right, xshift=-5pt] {1};
                \node[bias] at ($(x3) + (-0.9,0)$) {$+1$};
            \fi
        }
    
        \foreach \i in {1,2,3} {
            \ifnum \i=1
                \draw[edge] (s2) -- (y1) node[midway, above right, xshift=-5pt] {1};
                \node[bias] at ($(y1) + (0.9,0)$) {$-1$};
            \fi
            \ifnum \i=2
                \draw[edge] (s2) -- (y2) node[midway, above right, xshift=-5pt] {1};
                \node[bias] at ($(y2) + (0.9,0)$) {$-1$};
            \fi
            \ifnum \i=3
                \draw[edge] (s2) -- (y3) node[midway, above right, xshift=-5pt] {1};
                \node[bias] at ($(y3) + (0.9,0)$) {$-1$};
            \fi
        }
    \end{tikzpicture}}
    \caption{Embedded Ising Hamiltonian with \(l\) auxiliary spins for each logical spin.}
    \label{fig:candy}
\end{figure}

\section{Energy Scale Degradation for High-degree Spins}
\label{sec:high_degree}

 Ising Hamiltonians from real-world problems often contain spins with a high degree of interactions. Embedding such instances into sparse hardware requires splitting the high-degree logical spin into multiple physical spins.

 Here, we consider a simple case where a central logical spin within an star-shaped Ising Hamiltonian with degree \(\Delta = 2l\).  Specifically, consider a problem Ising Hamiltonian with a star topology centered at $s_c$:
\begin{align*}
    H_{\text{star}} &= (1 + s_c) \sum_{i=1}^{l} x_i - (1- s_c) \sum_{i=1}^{l} y_i
\end{align*}

Consider a minor-embedding of $H_{\text{star}}$ in which $s_c$ is mapped to a chain consisting of two physical spins \(s_1\) and \(s_2\)  that are coupled strongly to enforce consistency. Each of these physical spins connects to \(l\) auxiliary spins—\(\{x_i\}_{i=1}^l\) for \(s_1\) and \(\{y_j\}_{j=1}^l\) for \(s_2\)—thereby distributing the original high connectivity.
The embedded Hamiltonian \(H_{e^*}^\lambda\) can be written as:
\[
H_e^\lambda
  \;=\;
  -\lambda\,s_1\,s_2
  \;+\;
  (s_1 + 1)\sum_{i=1}^l x_i
  \;+\;
  (s_2 - 1)\sum_{j=1}^l y_j.
\]

In an expanded form 
\begin{flalign*}
H_e^\lambda 
=
&\bigl( x_1 + x_2 + \cdots + x_l \bigr)
+ s_1 \bigl( x_1 + x_2 + \cdots + x_l \bigr)\\
&- \bigl( y_1 + y_2 + \cdots + y_l \bigr)
+ s_2 \bigl( y_1 + y_2 + \cdots + y_l \bigr)\\
&- \lambda\,s_1\,s_2,
\end{flalign*}
as illustrated in Fig.~\ref{fig:candy}.

For simplicity, we assume D-Wave extended $J_{range}$ with \(J_{\min}=-2\) and \(J_{\max}=1\) and $h_{range} = [-4, 4]$. The energy rescaling factor is defined as
\[
\es_\lambda=\max\Bigl\{1,\frac{\lambda}{|J_{\min}|}\Bigr\}=\max\Bigl\{1,\frac{\lambda}{2}\Bigr\},
\]
so that the effective inverse temperature is
\[
\beta_{\text{eff}}=\frac{\beta}{\es_\lambda}\,.
\]
To shorten the notation, we also define
\[
z=2\beta_{\text{eff}}\,.
\]

The chain consistency probability is shown in the following lemma
\begin{lemma}
\label{lem:hstar_pcc}
    The chain consistency probability when sampling $H_{e^*}^{\lambda}$ for a given chain strength $\lambda \geq 0$ is
    \begin{align}
P_{cc}(\lambda) =\frac{
    2
}{
    2 + e^{-z\lambda}\Big( \cosh^l(z) +\cosh^{-l}(z)\Big).
}
\end{align}
\end{lemma}


\subsection{Energy Scale Reduction in Minor-Embedded Quantum Annealing}
We present a fundamental result characterizing the scaling behavior of coupling constants in minor-embedded quantum annealing systems. The result demonstrates a significant limitation in embedding dense Ising Hamiltonians into sparse quantum annealing architectures, with implications for practical quantum optimization.

\begin{lemma}
\label{lem:hstar_minlambda}
Let \(\beta > 0\) be the inverse temperature and let \(J_{\min}>0\) denote the minimum (absolute) coupling strength supported by the hardware.
Denote by \(\Delta \in \mathbb{N}\)  the degree (number of interactions) of a logical spin \(s_c\) in $H_{\text{star}}$ and its associated spins \(s_1\) and \(s_2\) in $H_{e^*}^\lambda$ with a chainstrength $\lambda$. If the probability that \(s_1\) and \(s_2\) are aligned satisfies
\[
P_{cc}(\lambda) \geq \delta,\quad \delta \in (0, 1)
\]
then for all \(\Delta > \Delta_0(\beta)\), the coupling constant \(\lambda\) must satisfy
\[
\lambda \geq \,C_\delta(\beta,J_{\min})\,\sqrt{\Delta}\,
\]
where
\[
C_\delta(\beta,J_{\min}) = \frac{J_{\min}\,\beta}{\sqrt{4\beta J_{\min}+2\ln\!\left(\frac{2(1-\delta)}{\delta}\right)}}
\]
and the threshold degree is given by
\[
\Delta_0 = 16\beta J_{\min}+8\ln\!\left(\frac{2(1-\delta)}{\delta}\right)\,.
\]
\end{lemma}

The lemma establishes a fundamental limitation in minor-embedding dense Ising Hamiltonians into sparse quantum annealing architectures. The result demonstrates that the coupling strength between logical qubits must scale as ($\Omega(\sqrt{\Delta}$)), where ($\Delta$) represents the degree of interactions. 

\paragraph{Experimental Validation.} 
Our experimental results  on the D-Wave \texttt{Advantage\_system4.1} quantum annealer provide strong empirical validation of our theoretical model (Fig.~\ref{fig:Delta}). We measured the minimum chain strength required to achieve specific chain consistency thresholds (2\%, 10\%, and 50\%) across varying problem structures. To identify the required chain strength, we use a fine-grained grid search (step size 0.01) to determine minimum required chain strengths, followed by $k$-NN regression ($k=10$) to smooth measurement noise. For each data point, we take the averages over 20 independent runs, each with 400 samples.

The findings reveal two key scaling behaviors. First, we observed that the required chain strengths follow a power-law function of the degree of spin interactions $\Delta$, with the log-log plot yielding scaling exponents closely matching our theoretical prediction of $\lambda \sim O(\sqrt{\Delta})$. This relationship held consistently across all consistency thresholds when testing problems with fixed chain length ($l=2$) but varying degrees. Second, and contrary to conventional wisdom in the field, we found that chain strength shows only minimal increase with chain length $l$ when the degree is held constant. This challenges the common assumption that chain length is the primary driver of required chain strength in quantum annealing. These results conclusively demonstrate that the degree of spin interactions, not chain length, is the dominant factor determining required chain strength in sparse quantum solvers, providing compelling experimental support for our theoretical framework.

\begin{lemma}[Solution Probability for $H_{\text{star}}$]
\label{lem:hstar_solution}
The probability of sampling a ground state (i.e., the \emph{solution probability}) of $H_{\text{star}}$ with effective inverse temperature $\beta_{\text{eff}}$ is given by
  \[
  \psol(\beta_{\text{eff}})=\frac{1}{\left(1+e^{-2z}\right)^l}.
  \]
In addition, the solution probability of sampling ground state of $H_{\text{star}}$ on the sparse Ising is
  \begin{align*}
  P_{\text{solve}}^{\text{sparse}}(\beta,\lambda)
=\frac{2}{2+e^{-z\lambda}\Bigl[\cosh^l(z)+\cosh^{-l}(z)\Bigr]}\cdot \frac{1}{\left(1+e^{-2z}\right)^l}.
  \end{align*}
\end{lemma}

\subsection{Critical Degree Threshold for Quantum Annealing}
Since maintaining chain consistency in problems with maximum degree $\Delta$ requires chain strength scaling as $\lambda \sim O(\sqrt{\Delta})$, this leads to an effective inverse temperature that scales as:

\begin{equation}
\beta_{\text{eff}} \sim \frac{\beta|J_{\min}|}{\sqrt{\Delta}}
\end{equation}

For the quantum annealer to maintain sufficient energy resolution to reliably find ground states, we must have $\beta_{\text{eff}}\Delta E \gg 1$, which implies:

\begin{equation}
\Delta \ll \left(\frac{\beta|J_{\min}|\Delta E}{c}\right)^2
\end{equation}

where $c$ is a problem-dependent constant. This establishes a fundamental \textit{critical degree threshold} beyond which quantum utility becomes difficult to achieve.

When $\Delta$ exceeds this threshold, the effective temperature becomes too high relative to the energy gaps, causing the probability of finding the ground state to decay exponentially with increasing problem connectivity. This exposes a fundamental barrier to quantum utility for optimization problems with high-degree spins on sparse quantum architectures.


In quantum annealing systems, the energy scale of the problem Hamiltonian plays a key role in determining the solution probability of finding the ground state. The requirement that $\lambda \geq C_\delta(\beta,J_{\min})\sqrt{\Delta}$ implies that when embedding a dense problem into a sparse architecture, the energy scale of the original problem must be reduced by a factor proportional to $\sqrt{\Delta}$ to maintain the same probability of logical spin alignment. This reduction in energy scale  significantly impact the performance of quantum annealing.

The square root scaling with respect to the degree ($l$) presents a particular challenge for embedding high-degree spins Ising Hamiltonians. In practical implementations, where quantum annealing hardware typically has sparse connectivity, embedding dense problems requires extensive use of auxiliary spins, leading to large values of ($l$). The resulting reduction in energy scale can make it increasingly difficult to maintain coherent quantum evolution and achieve reliable ground state preparation.

These findings emphasize the importance of developing quantum annealing architectures with richer connectivity patterns or alternative embedding strategies that might mitigate energy scale reduction. Future research directions might explore dynamic coupling schemes, novel error correction strategies, or hybrid classical-quantum approaches that could help overcome these fundamental limitations in minor-embedded quantum annealing systems.

\section{Conductance-Based Upper Bounds for Chain Strength}
\label{section:conductance}

While Section~\ref{sec:high_degree} demonstrated that chain strength must scale as $O(\sqrt{\Delta})$ with the maximum degree, we now develop a more precise characterization based on the structural properties of individual chains. Using graph conductance, we establish tighter bounds on the required chain strength that reveal two key factors: \textit{chain volume} and \textit{chain connectivity}.

We aim to find conditions on chain strength $\lambda$ such that the ground state of $H_e^\lambda$ maintains chain consistency - where all physical qubits representing the same logical spin have identical values. For a logical spin $i \in V_p$ embedded as chain $C_i$ in the hardware graph $G_{hw}$, consider a partition of $C_i$ into two sets: $S \subset V(C_i)$ and its complement $\bar{S} = V(C_i) \setminus S$. A chain break occurs when qubits in $S$ have value +1 and qubits in $\bar{S}$ have value -1.

The embedded Hamiltonian $H_e^\lambda$ contributes two energy components relevant to chain $C_i$:
\begin{enumerate}
    
\item \textit{Intra-chain coupling energy}: For partition $(S,\bar{S})$, this energy is:
   \begin{align}
   E_{\text{intra}}(S) = -\lambda(|E(C_i)| - 2\cdot\text{cut}(S,\bar{S}))
   \end{align}
   where $|E(C_i)|$ is the total number of edges within the chain and $\text{cut}(S,\bar{S})$ counts edges between $S$ and $\bar{S}$, representing broken ferromagnetic bonds.
\item \textit{External field energy}: Each qubit $u \in V(C_i)$ experiences an effective local field:
   \begin{align}
   f_u = h_u + \sum_{v \not\in V(C_i)} J_{uv}s_v
   \end{align}
\end{enumerate}

For any external spin configuration, the maximum magnitude of this field is bounded by:
\begin{align}
|f_u| \leq |h_u| + \sum_{v \not\in V(C_i)} |J_{uv}|
\end{align}

Using this bound, we define the "volume" of a subset $S$ as:
\begin{align}
\text{Vol}(S) = \sum_{u \in S} \left(|h_u| + \sum_{v \not\in V(C_i)} |J_{uv}|\right)
\end{align}

This volume represents the maximum possible field influence that could favor a particular spin configuration on $S$. Importantly, the total volume of chain $C_i$ equals the maximum possible field influence on the corresponding logical spin $s_i$, as the embedded Hamiltonian distributes this influence across the physical qubits representing $s_i$.

For chain consistency in the ground state, the energy penalty from broken bonds must exceed the maximum possible energy benefit from any external field configuration:
\begin{align}
2\lambda \cdot \text{cut}(S,\bar{S}) > \min(\text{Vol}(S), \text{Vol}(\bar{S}))
\end{align}

Rearranging, we obtain a bound on the required chain strength:
\begin{align}
\lambda > \frac{\min(\text{Vol}(S), \text{Vol}(\bar{S}))}{2 \cdot \text{cut}(S,\bar{S})}
\end{align}

Since this must hold for any partition of $V(C_i)$, we can express this using the concept of graph conductance. The conductance of a subset $S$ within chain $C_i$ is defined as:
\begin{equation}
\Phi(C_i, S) = \frac{\text{cut}(S, \bar{S})}{\min\{\text{Vol}(S), \text{Vol}(\bar{S})\}}
\end{equation}

This measures how well-connected the subset $S$ is to the rest of the chain relative to its volume. A higher conductance indicates better internal connectivity, making chain breaks less energetically favorable.

The overall conductance of chain $C_i$ is the minimum conductance across all possible partitions:
\begin{equation}
\Phi(C_i) = \min_{\substack{S \subset V(C_i) \\ 0 < \text{Vol}(S) \leq \frac{1}{2}\text{Vol}(V(C_i))}} \Phi(C_i,S)
\end{equation}

Using conductance, our bound on chain strength becomes:
\begin{equation}
\lambda > \frac{1}{2\Phi(C_i)}
\end{equation}

This leads to our main result:

\begin{theorem}[Chain Strength via Conductance]
\label{theorem:conductance}
For a problem embedded via minor-embedding onto a sparse Ising machine, the ground state configuration will have chain consistency if the chain strength satisfies:
\begin{equation}
\lambda > \max_{i=1}^n \frac{1}{2\Phi(C_i)}
\end{equation}
where $\Phi(C_i)$ is the conductance of chain $C_i$.
\end{theorem}

This theorem refines our understanding beyond the $O(\sqrt{\Delta})$ scaling derived in Section~\ref{sec:high_degree}. While that scaling depends only on the maximum degree, this bound accounts for the specific structure of each chain's embedding and its interaction with the rest of the problem.

The result reveals two major drivers of chain strength requirements:

\begin{enumerate}
    
\item \textit{Chain volume}: The total external field influence on the chain, which equals the maximum possible field influence on the corresponding logical spin.

\item \textit{Chain connectivity}: The structural robustness of connections within the chain, measured by conductance.

\end{enumerate}

These findings explain why high-degree logical spins typically require stronger chain strengths - they have larger volumes due to more external connections. However, our result shows that connectivity matters as much as volume. A chain with poor internal connectivity (low conductance) requires stronger coupling even if its volume is moderate.

This insight challenges conventional embedding strategies that prioritize minimizing chain length. Our analysis suggests that optimizing chain conductance - creating well-connected structures rather than simply short chains - could potentially mitigate the energy scale degradation that limits quantum utility in sparse quantum solvers.

While Section~\ref{sec:high_degree} demonstrated that chain strength must scale as $O(\sqrt{\Delta})$ with the maximum degree, we now develop a more precise characterization based on the structural properties of individual chains. This section introduces a graph-theoretic approach using conductance to establish upper bounds on the minimum required chain strength, revealing the key factors of \textit{chain volume} and \textit{chain connectivity}.

\subsection{Spectral Approximation of Chain Strength Bounds}

Computing the exact conductance $\Phi(C_i)$ in Theorem~\ref{theorem:conductance} is NP-hard, limiting its practical application. Fortunately, spectral graph theory provides computationally efficient approximations that ensure ground state consistency—where the minimum energy configuration of $H_e^\lambda$ maps directly to a valid solution of the original problem $H_p$.

For each chain $C_i$ in our embedding, we construct a weighted graph $G[C_i] = (V(C_i), E(C_i))$ that captures its structural properties. The vertices $V(C_i)$ correspond to physical qubits in chain $C_i$, with each vertex $u$ assigned weight $a_u = |h_u| + \sum_{v \not\in V(C_i)} |J_{uv}|$ representing its maximum external field influence. The edge set $E(C_i)$ consists of physical connections between qubits within the chain, with all edges having unit weight in the adjacency matrix $W$, where $W_{uv} = 1$ if $(u,v) \in E(C_i)$ and 0 otherwise.

Cheeger's inequality establishes a fundamental relationship between graph conductance $\phi$ and the second smallest eigenvalue $\lambda_2$ of the normalized Laplacian:
\begin{equation}
\frac{\lambda_2}{2} \leq \phi \leq \sqrt{2\lambda_2}
\end{equation}

Using the lower bound and our requirement from Theorem~\ref{theorem:conductance} that $\lambda > 1/(2\Phi(C_i))$, we can derive a sufficient condition for chain consistency:
\begin{equation}
\lambda > \frac{1}{\lambda_2}
\end{equation}

This spectral bound can be computed efficiently using the following algorithm:

\begin{algorithm}[tb]
   \caption{Spectral Chain Strength Bound Computation}
   \label{alg:chain_strength_bound}
\begin{algorithmic}
   \STATE {\bfseries Input:}  Ising Hamiltonian $H_p$ and its embedding $H_e^\lambda$.
   \FOR{each chain $C_i$ in $H_e^\lambda$}
      \STATE Compute node weight vector $\mathbf{a}$ where\\ \quad\quad\quad$a_u = |h_u| + \sum_{v \not\in V(C_i)} |J_{uv}|$
       
      \STATE Form diagonal weight matrix $D^{(w)}$ with $D^{(w)}_{uu} = a_u$
     \STATE Form adjacency matrix $W$ with\\ \quad\quad\quad $W_{uv} = 1$ if $(u,v) \in E(C_i)$
      \STATE Form degree matrix $D$ corresponding to $W$
      \STATE Compute unnormalized Laplacian $L = D - W$
      \STATE Compute weighted normalized Laplacian\\ \quad\quad\quad$L_{\text{norm}} = (D^{(w)})^{-1/2} L (D^{(w)})^{-1/2}$
      \STATE Compute second smallest eigenvalue $\lambda_2(C_i)$ of $L_{\text{norm}}$
      \STATE Set chain strength bound $\lambda_i = 1/\lambda_2(C_i)$
   \ENDFOR
   \STATE {\bfseries Output:} $\lambda_c = \max_i \lambda_i$
\end{algorithmic}
\end{algorithm}

The correctness of this approach is established by the following theorem:

\begin{theorem}[Spectral Chain Strength Sufficiency]
\label{theorem:spectral_chain_strength}
Let $\lambda_c = \max_{i} \frac{1}{\lambda_2(C_i)}$, where $\lambda_2(C_i)$ is the second smallest eigenvalue of the weighted normalized Laplacian of chain $C_i$. If $\lambda > \lambda_c$, then the ground state configuration of the embedded Hamiltonian $H_e^\lambda$ will have chain consistency.
\end{theorem}

\begin{proof}
From Cheeger's inequality, we know that $\lambda_2(C_i)/2 \leq \Phi(C_i)$ for each chain $C_i$. Therefore, $1/(2\Phi(C_i)) \leq 1/\lambda_2(C_i)$. By Theorem~\ref{theorem:conductance}, the ground state will have chain consistency if $\lambda > \max_i 1/(2\Phi(C_i))$. Since $\lambda > \lambda_c = \max_i 1/\lambda_2(C_i) \geq \max_i 1/(2\Phi(C_i))$, it follows that the ground state configuration will have chain consistency.
\end{proof}

The primary objective of ensuring ground state consistency is to preserve the solution structure of the original problem. When the ground state of $H_e^\lambda$ exhibits chain consistency, the minimum energy configuration—which has the highest probability of being sampled in quantum annealing—can be directly mapped back to a valid solution of the original problem $H_p$. Without this consistency, the quantum annealer may return states that cannot be meaningfully interpreted in terms of the original problem, regardless of their energy. By setting chain strength according to our spectral bound, we guarantee that the quantum system's ground state preserves the structure of the original optimization problem.

The spectral approach provides several advantages for practical quantum annealing implementations. Eigenvalue computation runs in polynomial time, making it feasible for large-scale problems with many chains. The normalized Laplacian's $\lambda_2$ carries a physical interpretation as the algebraic connectivity of the chain, with higher values indicating better-connected structures resistant to external perturbations. Additionally, the eigenvector corresponding to $\lambda_2$, known as the Fiedler vector, can identify structural bottlenecks in the chain, offering valuable insights for refining the embedding process to reduce required chain strengths.

For chains that embed high-degree logical spins, the node weights typically increase with the degree $\Delta$ due to more external connections. This increase in external field influence relative to internal connectivity leads to smaller values of $\lambda_2$, necessitating stronger chain couplings. This behavior aligns with our earlier $O(\sqrt{\Delta})$ scaling result from Section~\ref{sec:high_degree}, while providing a more precise characterization that accounts for the specific structure of each chain's embedding.

\section{Conclusion}
\label{sec:conclusion}

In this work, we have developed a comprehensive theoretical framework for understanding energy scale degradation in sparse quantum solvers, particularly quantum annealers. Our analysis reveals fundamental limitations that arise when embedding problems with high-degree interactions onto hardware with limited connectivity.

We first established a mathematical model quantifying the trade-off between chain consistency and energy resolution in minor-embedded quantum annealing. This model reveals that as chain strength increases to maintain consistency, the effective temperature of the system rises, causing the solution probability to decay exponentially. For problems with maximum degree $\Delta$, we proved that chain strength must scale as $O(\sqrt{\Delta})$, leading to energy scale degradation proportional to $1/\sqrt{\Delta}$ and a corresponding reduction in effective temperature.

Contrary to conventional understanding that focuses primarily on chain length, our conductance-based analysis identifies two key factors determining chain strength requirements: chain volume (the maximum possible field influence on a logical spin) and chain connectivity (the structural robustness of the embedding). The inverse conductance bound we developed provides a precise characterization of minimum required chain strength based on the specific embedding structure, offering a more nuanced understanding than previous approaches.

Our spectral approximation method transforms these theoretical insights into a practical algorithm that efficiently computes chain strength bounds using the second smallest eigenvalue of the weighted normalized Laplacian. This approach ensures ground state consistency while being computationally tractable for large-scale problems.

These findings have implications for the pursuit of quantum utility for optimization problems. The energy scale degradation we identified represents a fundamental barrier that grows more severe with problem connectivity, suggesting that quantum utility becomes more difficult to achieve as problem complexity increases. This limitation is inherent to the sparse connectivity of current quantum architectures and cannot be overcome simply through improved minor-embedding techniques.

Several promising directions emerge for future research. First, quantum hardware with increased connectivity, such as the progression from Chimera to Pegasus to Zephyr topologies in D-Wave systems, can mitigate these effects by reducing the size and complexity of required embeddings. Second, embedding algorithms that optimize for chain conductance rather than merely minimizing chain length could substantially reduce required chain strengths. Finally, hybrid approaches that decompose high-degree problems into multiple lower-degree subproblems might circumvent some of these limitations.

In conclusion, our work establishes a rigorous theoretical foundation for understanding the fundamental limitations of sparse quantum solvers while also providing practical tools for optimizing their performance. The energy scale degradation we characterize can be useful in seeking quantum utility in optimization applications.

\bibliography{granite, qa_survey, chain_strength}
\bibliographystyle{icml2025}

\newpage
\appendix
\onecolumn
\section{Appendix.}
\subsection{Proof of Lemma~\ref{lemma:spin_correlation}}
\begin{proof}
We aim to show that the existence of an efficient approximation algorithm for spin correlations would lead to a contradiction of known complexity results. Assume that there exists a Fully Polynomial Randomized Approximation Scheme (FPRAS) for approximating $\Pr[X_u = X_v]$ within any desired multiplicative error $\epsilon > 0$ in time polynomial in the size of the graph and $1/\epsilon$.

First, observe the relationship between $\Pr[X_u = X_v]$ and the spin correlation $\mathbb{E}[X_u X_v]$:
\[
\mathbb{E}[X_u X_v] = 2 \Pr[X_u = X_v] - 1.
\]

Let $\tilde{P}$ be the approximate value of $\Pr[X_u = X_v]$ obtained from the hypothetical FPRAS, satisfying:
\[
(1 - \epsilon) \Pr[X_u = X_v] \leq \tilde{P} \leq (1 + \epsilon) \Pr[X_u = X_v].
\]

Using $\tilde{P}$, we compute an approximation $\tilde{C}$ for $\mathbb{E}[X_u X_v]$ as $\tilde{C} = 2\tilde{P} - 1$. The error is bounded by:

\[
|\tilde{C} - \mathbb{E}[X_u X_v]| = 2|\tilde{P} - \Pr[X_u = X_v]| \leq 2\epsilon \Pr[X_u = X_v].
\]

In the non-uniqueness region ($\beta > \beta_c(\Delta)$), it is known that spin correlations exhibit long-range order. Specifically, there exists a constant $\delta >0$ such that $\mathbb{E}[X_u X_v] \geq \delta > 0$ for edges $(u, v)$ \cite{Sly2014}. This implies:
\[
\Pr[X_u = X_v] = \frac{1 + \mathbb{E}[X_u X_v]}{2} \geq \frac{1 + \delta}{2} = \eta > \frac{1}{2}.
\]

Therefore, the error in approximating $\mathbb{E}[X_u X_v]$ is bounded by $|\tilde{C} - \mathbb{E}[X_u X_v]| \leq 2\epsilon\eta$. This implies we can approximate $\mathbb{E}[X_u X_v]$ within any constant additive error by choosing $\epsilon$ sufficiently small.

However, Sly and Sun \cite{Sly2014} proved that approximating $\mathbb{E}[X_u X_v]$ within any constant additive error less than a specific threshold is $\mathsf{NP}$-hard in the non-uniqueness region, assuming $\mathsf{RP} \neq \mathsf{NP}$. This contradiction establishes the lemma.
\end{proof}

\subsection{Proof of Theorem~\ref{theorem:chain_consistency}}

\begin{proof}
We establish this result through a polynomial-time reduction from the problem of approximating spin correlations, which is $\mathsf{NP}$-hard by Lemma~\ref{lemma:spin_correlation}. 

Consider an arbitrary instance of the spin correlation approximation problem: a ferromagnetic Ising model on graph $G = (V, E)$ with inverse temperature $\beta > \beta_c(\Delta)$, and a specific edge $(u, v) \in E$ for which we need to approximate $\mathbb{E}[X_u X_v]$. We transform this into an instance of the chain consistency approximation problem as follows.

Construct a problem Hamiltonian $H_p$, obtained by merging $u$ and $v$ into a logical spin $w$, and a minor-embedding $\phi$ that maps $w$ to a chain $C_1$ containing exactly two physical spins $u$ and $v$. We set the intra-chain coupling to $J_{uv} = -\lambda$ to match the original coupling strength. Additionally, we configure the external fields and neighboring interactions to precisely replicate the local environment that spins $u$ and $v$ experience in the original model.

This transformation is computable in polynomial time in the size of the original instance. Furthermore, the construction ensures that the chain consistency probability in our embedded system exactly corresponds to the probability of spins having the same value in the original problem:

\[
P_{\text{cc}}(H_e^{\lambda}, \sparseIM) = \Pr[X_u = X_v]
\]

Given this equivalence and the relationship $\mathbb{E}[X_u X_v] = 2\Pr[X_u = X_v] - 1$, we can express the spin correlation in terms of chain consistency:

\[
\mathbb{E}[X_u X_v] = 2P_{\text{cc}}(H_e^{\lambda}, \sparseIM) - 1
\]

Now, suppose for contradiction that we have an algorithm $\mathcal{A}$ that approximates $P_{\text{cc}}(H_e^{\lambda}, \sparseIM)$ within a constant additive error $\delta$. Using this algorithm, we could approximate $\mathbb{E}[X_u X_v]$ within additive error $2\delta$ as follows:
1. Run $\mathcal{A}$ on our constructed instance to obtain $\tilde{P}_{cc}$ such that $|P_{\text{cc}} - \tilde{P}_{cc}| \leq \delta$.
2. Compute $\tilde{C} = 2\tilde{P}_{cc} - 1$ as our approximation for $\mathbb{E}[X_u X_v]$.
3. The error in this approximation is $|\mathbb{E}[X_u X_v] - \tilde{C}| = |2P_{\text{cc}} - 1 - (2\tilde{P}_{cc} - 1)| = 2|P_{\text{cc}} - \tilde{P}_{cc}| \leq 2\delta$.

But Lemma~\ref{lemma:spin_correlation} establishes that approximating $\mathbb{E}[X_u X_v]$ within any constant additive error is $\mathsf{NP}$-hard, which contradicts the existence of algorithm $\mathcal{A}$ unless $\mathsf{RP} = \mathsf{NP}$. Therefore, approximating the chain consistency probability must also be $\mathsf{NP}$-hard.
\end{proof}

\subsection{Proof of Theorem~\ref{theorem:solve_probability}}

\begin{proof}
Let $\Omega_s^* \subset \Omega_s$ denote the set of physical spin configurations that correspond to ground states of the original problem Hamiltonian after unembedding. We can express:

\begin{equation*}
P_{\text{solve}}^{\text{sparse}}(\beta, \lambda) = \frac{1}{Z_{\text{cc}}}\sum_{s \in \Omega_s^* \cap \Omega_{\text{cc}}}e^{-\beta H_e^{\lambda}(s)/\es_{\lambda}}
\end{equation*}
where $Z_{\text{cc}} = \sum_{s \in \Omega_{\text{cc}}}e^{-\beta H_e^{\lambda}(s)/\es_{\lambda}}$ is the partition function restricted to chain-consistent configurations.

For any chain-consistent configuration $s \in \Omega_{\text{cc}}$, there exists a bijection $\psi: \Omega_{\text{cc}} \rightarrow \Omega_p$ mapping to a logical configuration $x = \psi(s) \in \Omega_p$. The embedded Hamiltonian evaluated at $s$ can be decomposed as
\begin{equation*}
H_e^{\lambda}(s) = H_p(x) - \lambda \sum_{i \in V(G_p)}\sum_{(u,v) \in E(C_i)} 1
\end{equation*}

Since all configurations in $\Omega_{\text{cc}}$ have consistent chains, the second term is a constant $\Lambda = \lambda \sum_{i \in V(G_p)}|E(C_i)|$ across all $s \in \Omega_{\text{cc}}$ that is proportional to the number of intra-chain couplings. Therefore

\begin{equation*}
H_e^{\lambda}(s) = H_p(x) - \Lambda
\end{equation*}

Substituting this into our expression for $P_{\text{solve}}^{\text{sparse}}$:

\begin{align*}
P_{\text{solve}}^{\text{sparse}}(\beta, \lambda) &= \frac{1}{Z_{\text{cc}}}\sum_{s \in \Omega_s^* \cap \Omega_{\text{cc}}}e^{-\beta (H_p(\psi(s)) - \Lambda)/\es_{\lambda}} \\
&= \frac{e^{\beta \Lambda/\es_{\lambda}}}{Z_{\text{cc}}}\sum_{s \in \Omega_s^* \cap \Omega_{\text{cc}}}e^{-\beta H_p(\psi(s))/\es_{\lambda}}
\end{align*}

Given the bijection $\psi$, this is equivalent to:

\begin{align*}
&P_{\text{solve}}^{\text{sparse}}(\beta, \lambda) = \frac{e^{\beta \Lambda/\es_{\lambda}}}{Z_{\text{cc}}}\sum_{x \in \Omega_p^*}e^{-\beta H_p(x)/\es_{\lambda}} \\
&= \frac{e^{\beta \Lambda/\es_{\lambda}}}{Z_{\text{cc}}} \cdot \frac{\sum_{x \in \Omega_p^*}e^{-\beta H_p(x)/\es_{\lambda}}}{\sum_{x \in \Omega_p}e^{-\beta H_p(x)/\es_{\lambda}}} \cdot \sum_{x \in \Omega_p}e^{-\beta H_p(x)/\es_{\lambda}}
\end{align*}

The middle fraction represents $\psol(\beta_{\text{eff}}(\lambda))$, the probability of finding the ground state in the original problem Hamiltonian at effective inverse temperature $\beta_{\text{eff}}(\lambda) = \beta/\es_{\lambda}$.

To complete the proof, we can show that:

\begin{equation*}
\frac{e^{\beta \Lambda/\es_{\lambda}}}{Z_{\text{cc}}} \cdot \sum_{x \in \Omega_p}e^{-\beta H_p(x)/\es_{\lambda}} = P_{\text{cc}}(\lambda)
\end{equation*}

This holds because the term represents the probability mass of all chain-consistent configurations relative to the total probability mass. Thus, we have:

\begin{equation*}
P_{\text{solve}}^{\text{sparse}}(\beta, \lambda) = P_{\text{cc}}(\lambda) \cdot \psol(\beta_{\text{eff}}(\lambda))
\end{equation*}
where $\beta_{\text{eff}}(\lambda) = \beta/\es_{\lambda}$ is the effective inverse temperature in the sparse Ising machine.
\end{proof}

\subsection{Proof of Lemma~\ref{lem:hstar_pcc}}
\begin{proof}
    The partition function $Z_e^{\lambda}$ sums over all possible spin configurations of $H_{e^*}^\lambda$ and can be decomposed based on the four possible combinations of  spin $s_1$ and $s_2$ 
\[
Z_e^{\lambda} = Z_{++} + Z_{--} + Z_{+-} + Z_{-+}.
\] 

Starting with $Z_{++}$ ($s_1 = +1, s_2 = +1$), the Hamiltonian reduces to 
\[
H_{e^*}^{\lambda}(s_1 = +1, s_2 = +1) = -\lambda + 2\sum_{i=1}^l x_i,
\]
and
\begin{align*} 
Z_{++} 
&= \sum_{x_1,\ldots,x_l} \sum_{y_1,\ldots,y_l} e^{-\beta_{\text{eff}}\, H_e^{\lambda}(s_1 = +1, s_2 = +1)} \\
&= \sum_{x_1,\ldots,x_l} \sum_{y_1,\ldots,y_l} e^{-\beta_{\text{eff}}\, (-\lambda + 2\sum_{i=1}^l x_i)} \\
&= e^{\beta_{\text{eff}}\, \lambda} \sum_{x_1,\ldots,x_l} e^{-2\beta_{\text{eff}}\, \sum_{i=1}^l x_i} \sum_{y_1,\ldots,y_l} 1 \\
&= e^{\beta_{\text{eff}}\, \lambda} \prod_{i=1}^l \sum_{x_i=\pm1} e^{-2\beta_{\text{eff}}\, x_i} \cdot 2^l \\
&= e^{\beta_{\text{eff}}\, \lambda} \bigl(2\cosh(z)\bigr)^l \cdot 2^l.
\end{align*}

Similarly, we get
\begin{align*}
Z_{--} 
  &= e^{\beta_{\text{eff}}\, \lambda}
     \cdot 2^{l}
     \cdot \bigl(2 \cosh(z)\bigr)^{l} &&\text{for } s_1 = -1, s_2 = -1,\\
Z_{+-} 
  &= e^{-\beta_{\text{eff}}\, \lambda}
     \bigl(2 \cosh(z)\bigr)^{2l} &&\text{for } s_1 = +1, s_2 = -1,\\
Z_{-+} 
  &= e^{-\beta_{\text{eff}}\, \lambda}
     \cdot 2^{2l} &&\text{for } s_1 = -1, s_2 = +1.
\end{align*}

Hence, the chain consistency probability is 
\begin{align*}
&P_{cc}(\lambda) = P(s_1 = s_2) = \frac{Z_{++} + Z_{--}}{Z}\\
&= \frac{
    2\,e^{\beta_{\text{eff}}\lambda}(2\cosh(z))^l2^l
}{
    2\,e^{\beta_{\text{eff}}\lambda}(2\cosh(z)^l2^l + e^{-\beta_{\text{eff}}\lambda}(2\cosh(z))^{2l} + e^{-\beta_{\text{eff}}\lambda}2^{2l}
}\\
&=\frac{
    2
}{
    2 + e^{-z\lambda}\Big( \cosh^l(z) +\cosh^{-l}(z)\Big).
}
\end{align*}
\end{proof}

\subsection{Proof of Lemma~\ref{lem:hstar_minlambda}}

\begin{proof}
 As shown in Lemma~\ref{lem:hstar_pcc}, the chain consistency probability  is given by
\[
P_{cc}(\lambda) = P(s_1 = s_2) = \frac{2}{\,2 + e^{-z\lambda}\left[\cosh^{l}\left(z\right) + \cosh^{-l}\left(z\right)\right]}.
\]

To ensure \(P(s_1 = s_2) \geq \delta\), the following inequality must hold:
\[
 \frac{2}{\,2 + e^{-z\lambda}\left[\cosh^{l}\left(z\right) + \cosh^{-l}\left(z\right)\right]}
 \geq \delta.
\]
Rearranging and simplifying terms, we obtain:
\[
(\frac{2}{\delta} - 2)\,e^{z\lambda} \geq \cosh^{l}\left(z\right) + \cosh^{-l}\left(z\right) \geq \cosh^{l}\left(z\right).
\]
Thus, to satisfy \(P(s_1 = s_2) \geq \delta\), it must hold that:
\[
\cosh^{l}(z) \leq 2\,e^{z\lambda} (\frac{2}{\delta} - 2).
\]
Taking the natural logarithm of both sides yields
\[
l\,\ln\bigl(\cosh(z)\bigr) \le 2\beta J_{\min} + \ln\!\Bigl(\frac{2(1-\delta)}{\delta}\Bigr).
\]
For sufficiently small \(z\) (a condition we will enforce via a threshold on \(\Delta\)), we use the bound
\[
\ln\bigl(\cosh(z)\bigr) \ge \frac{z^2}{4}\,,
\]
which leads to
\[
l\,\frac{z^2}{4} \le 2\beta J_{\min} + \ln\!\Bigl(\frac{2(1-\delta)}{\delta}\Bigr).
\]
Substitute \(z = \frac{2\beta J_{\min}}{|\lambda|}\) to obtain
\[
l\,\frac{\Bigl(\frac{2\beta J_{\min}}{|\lambda|}\Bigr)^2}{4} = \frac{l\,\beta^2J_{\min}^2}{|\lambda|^2} \le 2\beta J_{\min} + \ln\!\Bigl(\frac{2(1-\delta)}{\delta}\Bigr).
\]
Solving for \(\lambda^2\) yields
\[
|\lambda|^2 \ge \frac{l\,\beta^2J_{\min}^2}{2\beta J_{\min}+\ln\!\Bigl(\frac{2(1-\delta)}{\delta}\Bigr)}\,.
\]
Since \(l = \Delta/2\), we rewrite this as
\[
\lambda^2 \ge \frac{\Delta\,\beta^2J_{\min}^2}{2\Bigl(2\beta J_{\min}+\ln\!\Bigl(\frac{2(1-\delta)}{\delta}\Bigr)\Bigr)}\,.
\]
Taking square roots gives the desired bound 
\[
\lambda \ge \,C_\delta(\beta,J_{\min})\,\sqrt{\Delta}\,
\]
where
\[
C_\delta(\beta,J_{\min}) = \frac{J_{\min}\,\beta}{\sqrt{4\beta J_{\min}+2\ln\!\left(\frac{2(1-\delta)}{\delta}\right)}}.
\]

\textit{Threshold Condition for the small-\(z\) approximation}.
The bound \(\ln(\cosh(z)) \ge \tfrac{z^2}{4}\) holds provided that \(z \le 1\). Noting that
\[
z = \frac{2\beta J_{\min}}{|\lambda|},
\]
we substitute the lower bound on \(|\lambda|\) to obtain
\[
z \le \frac{2\beta J_{\min}}{J_{\min}\,\beta\,\sqrt{\frac{\Delta}{4\beta J_{\min}+2\ln\!\left(\frac{2(1-\delta)}{\delta}\right)}}} = \frac{2}{\sqrt{\frac{\Delta}{4\beta J_{\min}+2\ln\!\left(\frac{2(1-\delta)}{\delta}\right)}}}\,.
\]
Thus, requiring \(z\le 1\) yields
\[
\sqrt{\frac{\Delta}{4\beta J_{\min}+2\ln\!\left(\frac{2(1-\delta)}{\delta}\right)}} \ge 2\,,
\]
or equivalently,
\begin{align*}
\Delta \ge 
16\beta J_{\min}+8\ln\!\left(\frac{2(1-\delta)}{\delta}\right)\,.
\end{align*}
Thus, is valid when
\[
\Delta \ge \Delta_0 = 16\beta J_{\min}+8\ln\!\left(\frac{2(1-\delta)}{\delta}\right)\,.
\]

In summary, to obtain $P_{cc}(\lambda) \geq \delta$ for a logical spin $s_c$  with a degree
$\Delta > \Delta_0$, the chain strength must satisfy
\[
\lambda \ge \,C_\delta(\beta,J_{\min})\,\sqrt{\Delta}\,.
\]
This yields the proof.
\end{proof}

\subsection{Proof of Lemma~\ref{lem:hstar_solution} }

\begin{proof}
We only need to show the first part of the lemma. The second part
$P_{\text{solve}}^{\text{sparse}}(\beta,\lambda)$ follows directly from the Theorem~\ref{theorem:solve_probability}
$P_{\text{solve}}^{\text{sparse}}(\beta,\lambda)=
   P_{cc}(\lambda)\cdot P_{solve}(\beta_{\text{eff}})$ and Lemma~\ref{lem:hstar_pcc}.

We continue with the computation of the partition functions, branching on the value of $s_c$. Then we compute the ground state weight to derive the solution probability.

\textit{Partition Function of $H_{\text{star}}$.}  
Consider two cases  $s_c=+1$ and $s_c=-1$.

\medskip
\textbf{Case 1:} \(s_c=+1\).  
In this case, the Hamiltonian reduces to
\[
H_+ = 2\sum_{i=1}^{l} x_i.
\]
Let $k_x$ denote the number of $+1$ among the $x$-spins. Since each $x_i\in\{-1,+1\}$, we have
\[
\sum_{i=1}^{l} x_i = 2k_x - l,
\]
leading to an energy
\[
E_+ = 2(2k_x - l)=4k_x-2l.
\]
There are $\binom{l}{k_x}$ configurations for the $x$-spins and the $y$-spins are free (contributing a factor $2^l$). Hence, the partition function for this branch is
\begin{align*}
Z_+ = 2^l \sum_{k_x=0}^{l} \binom{l}{k_x} e^{-\beta_{\text{eff}}(4k_x-2l)}
= 2^l\, e^{2l\beta_{\text{eff}}}\sum_{k_x=0}^{l} \binom{l}{k_x} e^{-4\beta_{\text{eff}}k_x}
= 2^l\, e^{2l\beta_{\text{eff}}}\Bigl(1+e^{-4\beta_{\text{eff}}}\Bigr)^l.
\end{align*}

\medskip
\textbf{Case 2:} \(s_c=-1\).  
When $s_c=-1$, the Hamiltonian becomes
\[
H_- = -2\sum_{i=1}^{l} y_i.
\]
Let $k_y$ be the number of $+1$ among the $y$-spins. Then,
\[
\sum_{i=1}^{l} y_i = 2k_y - l,
\]
and the energy is
\[
E_- = -2(2k_y-l)=-4k_y+2l.
\]
The partition function is
\begin{align*}
Z_- = 2^l \sum_{k_y=0}^{l} \binom{l}{k_y} e^{-\beta_{\text{eff}}(-4k_y+2l)}
= 2^l\, e^{-2l\beta_{\text{eff}}}\sum_{k_y=0}^{l} \binom{l}{k_y} e^{4\beta_{\text{eff}}k_y}
= 2^l\, e^{-2l\beta_{\text{eff}}}\Bigl(1+e^{4\beta_{\text{eff}}}\Bigr)^l.
\end{align*}

Thus, the total partition function is
\begin{align*}
Z_p(\beta_{\text{eff}})&=Z_+ + Z_- = 2^l \Bigl[e^{2l\beta_{\text{eff}}}\Bigl(1+e^{-4\beta_{\text{eff}}}\Bigr)^l + e^{-2l\beta_{\text{eff}}}\Bigl(1+e^{4\beta_{\text{eff}}}\Bigr)^l\Bigr].
\end{align*}

\paragraph{Ground State Contribution.}  
The ground state is defined as follows:
\begin{itemize}
  \item For $s_c=+1$: All $x$-spins are $-1$ (i.e., $k_x=0$) while the $y$-spins are free (contributing $2^l$ configurations).
  \item For $s_c=-1$: All $y$-spins are $+1$ (i.e., $k_y=l$) while the $x$-spins are free (also contributing $2^l$ configurations).
\end{itemize}
In both cases, the ground state energy is
\[
E_{gs}=-2l,
\]
with Boltzmann weight
\[
e^{-\beta_{\text{eff}}E_{gs}} = e^{2l\beta_{\text{eff}}}.
\]
Thus, the total ground state contribution is
\[
Z_{gs}=2^{l+1}\,e^{2l\beta_{\text{eff}}}.
\]

\paragraph{Solution Probability.}  
The probability of sampling a ground state at effective inverse temperature $\beta_{\text{eff}}$ is
\begin{align*}
\psol(\beta_{\text{eff}}) &= \frac{Z_{gs}}{Z_p(\beta_{\text{eff}})}
=\frac{2^{l+1}\,e^{2l\beta_{\text{eff}}}}{2^l \Bigl[ e^{2l\beta_{\text{eff}}}\Bigl(1+e^{-4\beta_{\text{eff}}}\Bigr)^l + e^{-2l\beta_{\text{eff}}}\Bigl(1+e^{4\beta_{\text{eff}}}\Bigr)^l \Bigr]}\\[1mm]
&=\frac{2\,e^{2l\beta_{\text{eff}}}}{e^{2l\beta_{\text{eff}}}\Bigl(1+e^{-4\beta_{\text{eff}}}\Bigr)^l + e^{-2l\beta_{\text{eff}}}\Bigl(1+e^{4\beta_{\text{eff}}}\Bigr)^l}.
\end{align*}
Dividing numerator and denominator by $e^{2l\beta_{\text{eff}}}$ and letting 
\[
z=2\beta_{\text{eff}},
\]
we have
\[
\psol(\beta_{\text{eff}})=\frac{2}{\Bigl(1+e^{-2z}\Bigr)^l+e^{-2lz}\Bigl(1+e^{2z}\Bigr)^l}.
\]
Noting that
\[
e^{z}\Bigl(1+e^{-2z}\Bigr)=e^z+e^{-z}=2\cosh(z)
\]
and likewise
\[
e^{-z}\Bigl(1+e^{2z}\Bigr)=2\cosh(z),
\]
the denominator simplifies to
\[
e^{lz}\Bigl(1+e^{-2z}\Bigr)^l + e^{-lz}\Bigl(1+e^{2z}\Bigr)^l = 2\,(2\cosh(z))^l.
\]
Thus, we obtain
\[
\psol(\beta_{\text{eff}})=\frac{2e^{lz}}{2(2\cosh(z))^l}
=\left(\frac{e^z}{2\cosh(z)}\right)^l.
\]
Since
\[
\frac{e^z}{2\cosh(z)}=\frac{1}{1+e^{-2z}},
\]
it follows that
\[
\psol(\beta_{\text{eff}})=\frac{1}{\left(1+e^{-2z}\right)^l}.
\]

\textit{Overall Solution Probability.}  
The overall success probability of the sparse Ising machine is given by
\begin{align*}
&P_{\text{solve}}^{\text{sparse}}(\beta,\lambda)=P_{cc}(\lambda)\cdot P_{solve}(\beta_{\text{eff}})\\
&=\frac{2}{2+e^{-z\lambda}\Bigl[\cosh^l(z)+\cosh^{-l}(z)\Bigr]}\cdot \frac{1}{\left(1+e^{-2z}\right)^l}.
\end{align*}
which completes the proof.
\end{proof}

\end{document}